\documentclass[preprint]{aastex}

\newbox\grsign \setbox\grsign=\hbox{$>$} 
\newdimen\grdimen \grdimen=\ht\grsign
\newbox\laxbox \newbox\gaxbox
\setbox\gaxbox=\hbox{\raise.5ex\hbox{$>$}\llap
     {\lower.5ex\hbox{$\sim$}}}\ht1=\grdimen\dp1=0pt
\setbox\laxbox=\hbox{\raise.5ex\hbox{$<$}\llap
     {\lower.5ex\hbox{$\sim$}}}\ht2=\grdimen\dp2=0pt
\def\gax{\mathrel{\copy\gaxbox}}
\def\lax{\mathrel{\copy\laxbox}}

\begin{document}

\title{
X-ray Spectral Diagnostics of Gamma-Ray Burst Environments
}
\author{Frits Paerels \altaffilmark{1}
Erik Kuulkers, John Heise}
\affil{SRON Laboratory for Space Research, \\
Sorbonnelaan 2, 3584 CA Utrecht, the Netherlands}
\and
\author{Duane A. Liedahl}
\affil{Department of Physics and Space Technology, 
Lawrence Livermore National Laboratory, \\
P.O. Box 808, L-41, Livermore, CA 94550, USA}
\altaffiltext{1}
{present address: Columbia Astrophysics Laboratory,
Columbia University, 538 W. 120th St., New York, NY 10027, USA}

\begin{abstract}

Recently, the detection of discrete features 
in the X-ray afterglow spectra of GRB970508 and GRB970828 was
reported. The most natural interpretation of these features is that
they are redshifted Fe K emission complexes. 
The identification of the line emission mechanism has drastic
implications for the inferred mass of radiating material, and hence
the nature of the burst site. X-ray spectroscopy provides a direct
observational constraint on these properties of gamma-ray bursters.
We briefly discuss how these constraints arise, in the context of an
application to the spectrum of GRB970508.

\end{abstract}


\section{Introduction}

The detection of a discrete spectral feature in the X-ray
afterglow spectrum of GRB970508 was recently 
reported by Piro {\it et al.}
(1999a,b). A similar feature in the afterglow of GRB970828 was 
reported by Yoshida {\it et al.} (1999). The redshift of the host
galaxy of GRB970508 was determined to be $z = 0.835$ (Metzger {\it et
al.} 1997; Bloom {\it et al.} 1998). 
The apparent energy of the discrete feature in the X-ray
spectrum is consistent with the redshifted energies of the $n = 2-1$
transitions in all possible charge states of Fe, {\it i.e.}, from Fe
K$\alpha$ at 6.4 keV, to H-like Fe Ly$\alpha$ at 6.95 keV. Therefore,
none of the possible line excitation mechanisms (fluorescence,
collisional excitation in hot gas, or
recombination in photoionized gas or a transient collisional plasma) 
is currently ruled out on the basis
of the measured line energy alone. 

The conditions required for each of these emission mechanisms ({\it
i.e.} the presence of a given charge state of Fe) are all still
compatible with the constraints on ionization parameter and gas
temperature that can be derived from the fact that the line source
must be located no more than about 1 light day from the ionizing
source, and from the shape of the ionizing spectrum. All three
mechanisms require high gas density ($n_e \gax 10^{11}$ cm$^{-3}$) for
bound Fe to exist at all this close to the burst site. 
The presence of a large mass of dense
gas close to the burst
has led to the suggestion that
a merging neutron star-neutron star binary is an unlikely site for
this burst, and that a scenario involving some sort of stellar
collapse is more likely (Piro {\it et al.} 1999b). 
In that case, the line source
may be associated with debris from the stellar collapse. 

Resolved X-ray spectroscopy can of course distinguish between the 
various line emission mechanisms, and lead to the correct
characterization of the conditions in the source and the implied mass
of radiating Fe. In the following, we
will briefly recapitulate the constraints imposed by proximity and the
energy budget, and then draw attention to an example of direct spectroscopic
analysis of the X-ray spectrum:
the fact 
that the measured centroid
position, and to some extent even the shape, of the X-ray emission 
feature in the afterglow of GRB970508 already imply
that recombination in afterglow-photoionized gas is unlikely to be the
emission mechanism for this burst. 

Our analysis applies to an optically thin, homogeneous medium. 
Complications arising from radiative transfer effects or external
heat input by a relativistic shock may conspire to alter the shape of
the spectrum, perhaps to the point that no definitive conclusions can be drawn
from an intantaneous, low sensitivity spectrum. Such effects have been
calculated by Weth et al. (1999) for the case of photoionization in
the context of two specific geometries for the burst site, 
and by Vietri (1999) for a
scenario in which the relativistic shock heats the medium shortly
after the onset of the afterglow. With better data, it may 
hopefully be possible to reverse the argument, and infer the geometry,
thermal history, and abundances from spectroscopy.

\section{Ionization and Thermal Conditions in the GRB970508 X-ray Line
Source}

The discrete feature appears in the {\it BeppoSAX} Medium Energy
Concentrator Spectrometer (MECS) spectrum immediately before a sudden
rise in the afterglow flux, beginning at $\sim 6 \times 10^4$ sec
after the burst, and disappears thereafter (Piro {\it et al.} 1999b).
If one attributes the feature to emission by Fe, one requires
bound Fe to be present close to the source, which emits a very large
flux of ionizing photons in the afterglow
($L_{1\ {\rm Ryd}-10\ {\rm keV}} \sim 10^{51} (t/1\
{\rm sec})^{-1.1}$ erg s$^{-1}$;
here, $t$ is the time
since the burst [Piro {\it et al.} 1999b]). 

The prompt burst and afterglow are intense enough that the
photoionization timescale, $t_{\rm ion}$, is extremely short compared
to the burst duration and the time during which the emission feature
is visible. Specifically, the 
(inverse of the) ionization timescale for Fe in the
afterglow radiation field is
\begin{eqnarray}
t_{\rm ion}^{-1} & = & \int_{\chi}^{\infty} dE\ F(E)\cdot E^{-1} 
\sigma(E) = \\
\nonumber
& = & 1.4 \times 10^4\ 
(t/{\rm 1\ sec})^{-1.1}\ r_{16}^{-2}\ {\rm sec^{-1}},
\end{eqnarray}
with $F(E)$ the ionizing flux at the line source, 
$\sigma(E) \approx 6 \times 10^{-18} 
\ Z^{-2}(\chi/E)^3$ cm$^2$ 
the K-shell photoionization cross section for neutral
atoms of nuclear charge $Z$, $\chi$ the ionization potential, and 
$r_{16}$ the distance to the burst site in units $10^{16}$ cm. We
have assumed a $E^{-2}$ photon number 
spectrum to calculate $F(E)$ from the $2-10$ keV
afterglow flux quoted by Piro {\it et al.} (1998), converted to
luminosity assuming 
$H_0 = 75\ {\rm km}\ {\rm s}^{-1} {\rm Mpc}^{-1}$ and
$q_0 = 1/2$, and we extrapolate the luminosity to 1 MeV.
Ionization is essentially instantaneous.

The recombination timescale is given by $t_{\rm rec} \sim 
(n_e \alpha(T_e))^{-1}$, where $\alpha(T_e)$ is the recombination
coefficient, $n_e$ the electron density,
and $T_e$ the electron temperature. Since $\alpha$ depends
on the temperature, we first 
need to consider the thermal evolution of the
source subject to irradiation by the prompt burst and afterglow
radiation fields.

Given sufficient
interaction time, the gas will relax to the Compton
temperature, $T_{\rm C}$, at which the Compton heating and cooling
rates equal each other, and which is therefore 
determined only by the shape of the
ionizing spectrum. 
In the non-relativistic limit (both the photon energies $E$ and the
electron energies $kT_e \ll m_e c^2$), the Compton temperature is
given by 
\begin{equation}
kT_{\rm C} = {1 \over 4} {{\int_{}^{} dE\ E\ F(E)}\over{\int dE\ F(E)}}
\end{equation}
(Ross 1979).
As either the photon energies or the gas temperature approach the
electron rest energy, the correct relativistic scattering cross
section should be used, and the energy exchange between photons and
electrons should be calculated to higher order than linear in 
$E/m_e c^2$ and $kT_e/m_e c^2$. In
order to obtain a very rough estimate, we will assume here that we can
mimic the effects of these modifications by simply cutting off the
integrals at photon energies of order 1 MeV. If we assume a prompt
burst spectrum of the form $F(E) \propto E^{1/2}$ at low energies, and
$F(E) \propto E^{-3/2}$ above a break energy of order 1 MeV
(Conners {\it et al.} 1998), the
Compton temperature becomes approximately $kT_{\rm C}\ {\rm (MeV)} \sim
(3/16)(E_{\rm max}/1\ {\rm MeV})^{1/2}$, with $E_{\rm max}$
the cutoff energy. For $E_{\rm max}$ in the few MeV range, the Compton
temperature is therefore in the several hundred keV range. 
The timescale for
the Compton interactions to equilibrate should be of order 
(Rybicki \& Lightman 1979)
\begin{equation}
t_{\rm Compton} = {{1}\over{n_e \sigma_{\rm T} c}}
{{m_ec^2}\over{4kT_e}} 
\sim 7 \times 10^4 n_{10}^{-1} T_8^{-1}\ {\rm sec}
\end{equation}
with $\sigma_{\rm T}$ the Thomson cross section. 
Unless the density is much higher than $10^{10}$ cm$^{-3}$,
this timescale is 
much longer than the burst timescale, which implies that the Compton
temperature will not be reached. Moreover, the 
average photon energy in the
burst spectrum is of order 1 MeV, which produces mildly relativistic
electrons. The stopping timescale of these fast particles on
stationary electrons is ({\it e.g.} Longair 1992)
\begin{equation}
t_{\rm e-e} \sim {{\gamma m_e^2 c^3}
\over{2\pi e^4 n_e \ln\Lambda}} 
= 1.3 \times 10^3 n_{10}^{-1}\ (\gamma/2)\ ((\ln \Lambda)/10)^{-1}
\ {\rm sec},
\end{equation}
(with $\gamma$ the electron Lorentz factor, and $\ln\Lambda$ the
Coulomb logarithm)
which leads to the interesting conclusion that the plasma may not be
in equilibrium at the end of the burst.

In any case, as 
the (much softer) afterglow begins, the plasma will rapidly cool by
inverse Compton scattering, 
bremsstrahlung, and, at low temperatures, atomic
emission.
For an initial gas temperature of order $10^8$ K or hotter, the
Comptonization timescale is of the order of, or shorter than the
duration of the afterglow, and equilibrium is established at the
afterglow 
Compton temperature, which for an $E^{-2}$ photon spectrum is equal to
$kT_{\rm C} = (1/4)E_{\rm max}/
\ln(E_{\rm max}/E_{\rm min})$, with 
$E_{\rm min}$ the lowest photon energy in the afterglow spectrum;
$E_{\rm max}$ is the upper end of the integration range, $E_{\rm max}
\sim$ 1 MeV.
Assuming the afterglow spectrum flattens below $E_{\rm min} \lax 1$ Ryd
(Wijers \& Galama 1999),
we find $T_{\rm C} \lax 2 \times 10^8$ K. 
Kallman and McCray (1979) in their numerical models for 
X-ray photoionized nebulae, actually find that the electron temperature
saturates at a lower temperature of approximately $T_{\rm C} \sim 
10^7$ K, due to bremsstrahlung cooling,
so most likely the gas is cooler than
$10^8$ K in the afterglow.

The recombination coefficient as a function of temperature 
for recombination onto a bare nucleus 
is given by Seaton (1959); a powerlaw fit 
to the coefficient is $\alpha \approx 1.26 \times 10^{-6} T_e^{-0.75}$
cm$^3$ s$^{-1}$ (for Z = 26),
accurate to $\lax 20\%$ in the range $T_e = 10^5 - 2 \times 10^8$ K.
Thus, the recombination
timescale is
\begin{equation}
t_{\rm rec} = 1.4\ n_{11}^{-1}\ T_7^{0.75}\ {\rm sec}.
\end{equation}
We conclude that at sufficiently high densities 
($n_{11} \gax 1$), the
line source will be relatively cool, and close to ionization
equilibrium at all times during the afterglow.

Assuming ionization equilibrium, we can derive a
constraint on the density from the 
requirement that bound Fe be present in the line source. 
For Fe to recombine to
the H-like charge state requires an ionization
parameter in equilibrium of $\xi \equiv L/n r^2 \lax 10^4$, with $L$
the ionizing luminosity, $n$ the particle density, and $r$ the
distance to the ionizing source. This value of the ionization
parameter applies to an $E^{-2}$ ionizing photon spectrum
(Kallman \& McCray 1982, their
model 7). Inserting numbers and extrapolating the ionizing luminosity
to 1 MeV,
one infers a lower limit to the density
at a distance of $6 \times 10^4$ lt sec from the source, of
\begin{equation}
n \gax 1.7 \times 10^{11}\ (t/{6\times 10^4\ {\rm sec}})^{-1.1} 
(r/6 \times 10^4\ {\rm lt\ sec})^{-2}\ {\rm cm}^{-3}
\end{equation}
Similarly, for Fe to be recombined to Li-like or less ionized,
necessary for fluorescence to operate, $\xi \lax 1000$
in equilibrium, yielding 
a ten times higher density than the above estimate.

\section{X-ray Spectroscopy}

From the previous we conclude that 
if the interaction with the afterglow radiation is the only source of
heating in the line emitting gas, then, later in the afterglow, 
the source is likely to be dense
($\gax 10^{11}$ cm$^{-3}$) and relatively cool.
Line emission can be excited both by fluorescence and
cascading following recombination. Only if there is an additional
source of heat, sufficient to heat the gas to $T \gax 10^8$ K, 
will collisional excitation play a role in driving Fe $n=2-1$ lines.

Obviously, X-ray spectroscopy of the discrete emission strongly
distinguishes between these various possible emission mechanisms. 
If one is confident of the redshift of the emission complex,
fluorescence can be distinguished from emission by highly
ionized Fe, simply from the apparent energy of the emission. 
Unfortunately, the resolution of the MECS, and the statistical quality
of the spectra under discussion are not sufficient to allow this
distinction to be made. 

Recombination and collisional excitation, however, can readily be
distinguished in the following way. For a given degree of ionization,
a photoionized medium in equilibrium is much cooler than the
corresponding collisional plasma, in which the electron temperature
has to be of order the ionization potential in order to support a
given charge state. The low electron temperature in photoionization
equilibrium has a dramatic spectroscopic signature. Since most of the
free electrons have small kinetic energies compared to the ionization
potential, photons resulting from radiative recombination will have a
narrow energy distribution, bunched up just above the ionization
potential, with a typical width of order $\Delta E \sim kT_e \ll \chi$.
In a collisional plasma on the other hand 
($kT_e \sim \chi$), the recombination rate is
much reduced, and in addition, the recombination photons are spread
out over a wide energy range above the ionization potential. The
narrow radiative recombination continuum (RRC) in a photoionized
source is easy to detect, because it contains an integrated photon
flux comparable to that in the $n=2-1$ discrete transition; in fact,
for H-like Fe, the ratio between the fluxes in the RRC to that in the
Ly$\alpha$ line is approximately equal to $0.93\ (kT_e/1\ {\rm
keV})^{0.21}$. 
So far, narrow RRC's have only been seen 
in the spectrum of the massive binary Cyg
X-3 (Liedahl \& Paerels 1996; Kawashima \& Kitamoto 1996). 

\section{Application to the Spectrum of GRB980508}

We use the dataset shown by Piro {\it et al.} (1999b; their dataset 1a).
At the resolution of the MECS ($\Delta E \sim 340$ eV at 3 keV), the
RRC is barely resolvable if the electron temperature is of order or
less than 1 keV. For simplicity we 
use a model for the emission spectrum of
cool, photoionized gas consisting of a narrow line plus exponentially
decaying RRC, to represent 
the Ly$\alpha$ emission line and RRC at 6.95 and 9.28 keV,
respectively, with equal photon fluxes.
The He-like recombination spectrum would look much the same at the
MECS resolution. To this, we added a simple power law with absorption
by neutral gas, to represent the continuum. We let the
redshift float, as well as the continuum parameters.

By fixing the electron temperature ({\it i.e.}, the width of the RRC)
and fitting for the remaining free parameters we obtain the minimum
$\chi^2$ values displayed in Figure 1. 
At either very low $T_e$ ($kT_e
\lax 0.2$ keV) or very high $T_e$ ($kT_e \gax 4$ keV) we obtain
$\chi^2$ values of order $\chi_{\rm min}^2 \approx 7$ for 5 degrees of
freedom. In themselves 
these values are acceptable (given the small number of
degrees of freedom). However, at low temperatures the model has a best
fitting redshift of $z > 1.6$ (for Hydrogenic Fe; 68\% confidence for
one parameter of interest only).
At high temperatures, the RRC becomes essentially undetectable and the
model consists effectively of a single narrow line. This model will
of course fit the data, but at these temperatures the plasma would 
be closer to collisional equilibrium, which is not consistent with the
assumption we set out to test. Finally, at $kT_e \sim 1$ keV (which is
about the temperature for a source in photoionization
equilibrium with the afterglow), we find $\chi_{\rm min}^2 \approx 12$,
which indicates a poor fit ($\chi_{\rm min}^2$ about two standard
deviations away from the expected $\chi_{\rm min}^2 = 5$). By itself,
this is probably not enough to confidently reject the model, and the
fit can be improved by changing the temperature somewhat. But
the implied redshift is still $z \sim 1.20$,
higher than for a single line because we have an additional emission
component at $4/3$ times the line energy. Unless we
are willing to entertain the possibility that the line source is
actually behind the galaxy at $z = 0.835$, we conclude that the
emission spectrum from a cool photoionized source is incompatible
with the measured spectrum. In Figure 1, we have indicated the
(68\% confidence) 
limits on the redshift as a function of the assumed temperature of the
gas, assuming H-like emission. At high temperature, the contrast in
the RRC becomes very small, and the spectrum is dominated by a single
line. In this limiting case, the implied redshift approaches $\sim
1.05 \pm 0.05$ (68\% confidence), appropriate for H-like Fe Ly$\alpha$. 
Widening the confidence interval to 90\% confidence will produce
agreement with the optical redshift of $z=0.835$ (Piro {\it et al.}
1999b).

Figure 2 displays the data and 
the lowest$-\chi^2$ recombination model with $kT_e = 1$ keV; the 
values for the other spectral parameters are: power law photon index
$-3.1$, column density $N_H = 7.1 \times 10^{21}$ cm$^{-2}$. This
continuum spectral shape appears to be quite a bit steeper than the
best fit given by Piro \& al. 1999b, but in fact, the uncertainty on
the index is large ($\approx 1.1$ for one parameter of interest, at
1$\sigma$), and so the continuum shapes are in fact consistent.

\section{Conclusion}

We find that Fe emission from a cool photoionized source in equilibrium
with the afterglow cannot account for the X-ray spectrum of GRB970508.
Fitting a spectral model for such a source either implies a large
redshift, or indicates such high electron temperatures that the source
would be closer to collisional ionization equilibrium. 

This would imply that the discrete feature in the spectrum of
GRB970508 arises from either Fe fluorescence or from collisional
excitation, each of which implies
a very different estimate for the total
mass of radiating Fe
($M_{\rm Fe} \sim 2.6 \times 10^{-4} M_{\odot}$ for fluorescence, 
$M_{\rm Fe} \sim 7 \times 10^{-2} M_{\odot}$ for collisional
excitation, which corresponds to 
$M_{\rm total} \sim 8 \times 10^{-2} M_{\odot}\ 
(A_{\rm Fe}/A_{{\rm Fe},\odot})^{-1}$ and
$M_{\rm total} \sim 28 M_{\odot}\ 
(A_{\rm Fe}/A_{{\rm Fe},\odot})^{-1}$, respectively (see, for
instance, Meszaros \& Rees [1998], 
Piro et al. [1999b], Lazzati et al. [1999]); here, 
$A_{\rm Fe}$ is the abundance of Fe).

With higher spectral resolution and sensitivity, we may be able to
distinguish spectroscopically between a fluorescent and a collisional
spectrum (a recombination spectrum is readily identified from the
presence of the RRC). Working from the K-shell spectrum 
of a single ionization stage of Fe alone, one
might use the $n=2-1$ and $n=3-1$ transitions, which are resolved at 
$E/\Delta E \sim 10$. Intensity ratios do not effectively
discriminate between collisional excitation and fluorescence:
$I({\rm Ly}\beta)/I({\rm Ly}\alpha) = 0.10-0.15$ (Hydrogenic Fe)
for electron
temperatures between $3 \times 10^7$ and $10^9$ K (Mewe, Gronenschild,
\& van den Oord 1985), while for neutral Fe 
$I({\rm K}\beta)/I({\rm K}\alpha) =
0.13$ (Kaastra \& Mewe 1993). Instead, with somewhat higher
resolution, one might use the fact that the ratio of the transition 
energies depends on the charge state: 
$E({\rm Ly}\beta)/E({\rm Ly}\alpha) = 1.18$, while $E({\rm
K}\beta)/E({\rm K}\alpha) = 1.10$.

At higher resolution, one would resolve the $1s-2p_{1/2}$ and
$1s-2p_{3/2}$ transitions ($E/\Delta E > 500)$. The ratios of the
energies are $E({\rm Ly}\alpha_2)/E({\rm Ly}\alpha_1) = 1.0030$ and
$E({\rm K}\alpha_2)/E({\rm K}\alpha_1) = 1.0020$, 
and a resolving power of
1000 is required to distinguish between these two. 

A collisional source at $kT < 15 $ keV will also show emission from
He-like Fe, and the simultaneous appearance of both the H- and the
He-like charge states readily indicates a high degree of ionization of
the source.
For temperatures in the range $3 < kT < 15$ keV, 
the He-like emission dominates,
which could still be identified as such if the characteristic
"triplet" structure of the $n=2-1$ lines is resolved ($E/\Delta E >
300$). The highest resolution available at energies above 2 keV is provided
by the High Energy Transmission Grating Spectrometer on {\it Chandra} 
(Canizares et al. 2000). The Fe K spectrum will be resolved at $E/\Delta E 
\approx 150~(1+z)$, which would at least enable charge-state spectroscopy at
moderately large redshift.

Finally, given sufficient sensitivity at low energies (and
sufficiently low absorption in the source), we may of course
detect emission from lower-$Z$ elements, either collisional or
fluorescent, which would greatly facilitate the spectroscopic
diagnosis of the excitation mechanism.
With the grating spectrometers on {\it Chandra} and {\it XMM}, 
standard detailed plasma diagnostics can be invoked, at resolving powers 
of $E/\Delta E \sim 100-1500$ at $E \lax 2$ keV,
depending on photon energy (Canizares et al.
2000, Brinkman et al. 2000, Brinkman et al. 1998).
Clearly, it would be very interesting to
pursue more sensitive X-ray spectroscopy of gamma ray burst afterglows.

\vskip 0.1in

We gratefully acknowledge discussions with Marten van Kerkwijk.
D. A. L. was supported in part by a
NASA Long Term Space Astrophysics Program grant (LTSA
S-92654-F).  Work at LLNL was performed under the
auspices of the U. S. Department of Energy, Contract No.
W-7405-Eng-48.

%\clearpage

\newpage

\noindent
Figure Captions:

\figcaption{
Minimum $\chi^2$ and best fitting redshift 
for fitting a spectral model appropriate for emission
from afterglow-photoionized gas to the {\it BeppoSAX}
MECS+LECS spectrum of GRB970508, as a function of electron
temperature $kT_e$. The grey band represents the 1$\sigma$ confidence
volume for the X-ray redshift; the dotted lines indicate the 1$\sigma$
contours for
the best-fitting X-ray redshift assuming 
H-like Ly$\alpha$ line emission only (no RRC).
}

\figcaption{
Best fitting photoionization equilibrium emission model, for an
assumed electron temperature of $kT_e = 1$ keV. Solid triangular
datapoints are LECS fluxes, solid circles are MECS fluxes. This model
implies a source redshift of $z = 1.20$ for H-like Fe emission.
Lower panel displays the post-fit residuals.
}

\vfill\eject

\centerline{\null}
\vskip7.5truein
\includegraphics{fig1.ps}

\vfill\eject

\centerline{\null}
\vskip7.5truein
\includegraphics{fig2.ps}

\vfill\eject

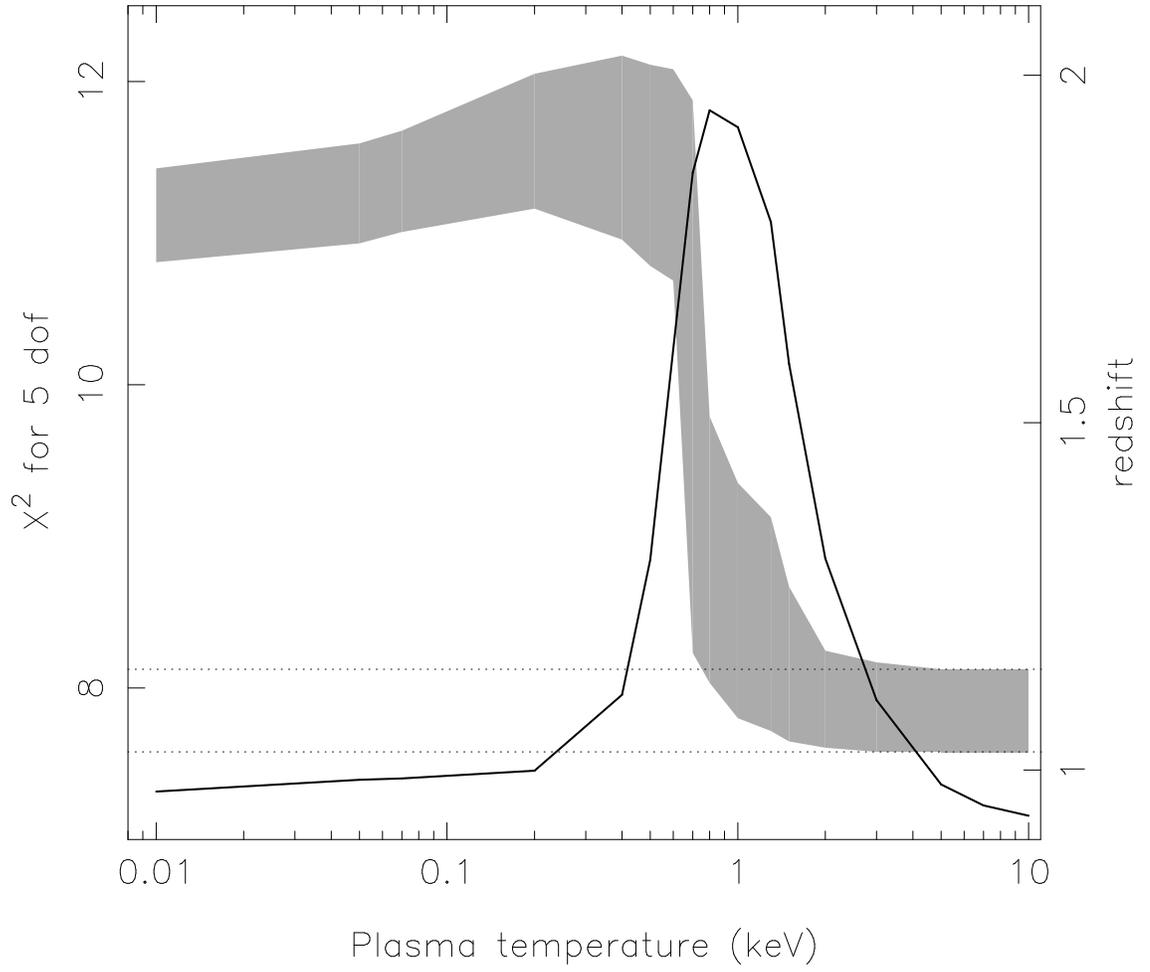
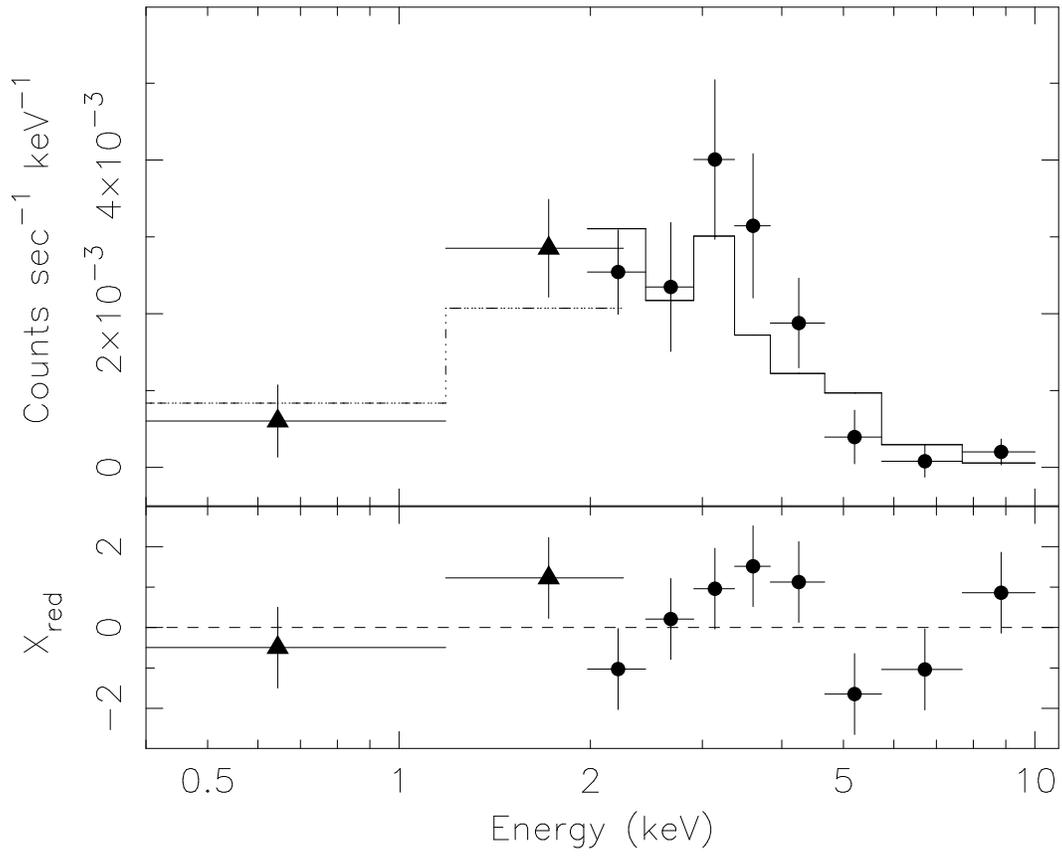
\end{document}